\documentstyle[prb,aps]{revtex}

\begin{document}
\draft
\title{Exact solution of the open Heisenberg chain with two impurities}
\author{Yupeng Wang}
\address{Cryogenic Laboratory, Chinese Academy of Sciences, Beijing 100080, 
P. R. China}
\maketitle
\begin{abstract}
We propose an integrable model of the spin-1/2 Heisenberg chain coupled to two 
impurity moments. With the open boundary conditions at the impurity sites, the
 model can be exactly solved for arbitrary impurity spin and arbitrary exchange 
 constants between the bulk and the impurities. The absence of redundant terms 
 in the hamiltonian makes the model very reasonable. The hamiltonian is 
 diagonalized via algebraic Bethe ansatz. It is found that the impurity spins 
 can only be screened (partially for $S>1/2$) for antiferromagnetic coupling 
 between the impurity  and the bulk. Otherwise the impurity spins can not be 
 screened. The residual entropy of the ground state and the Kondo temperature 
 are also derived explicitly based on the thermodynamic Bethe ansatz and the 
 local Fermi liquid theory.
\end{abstract}
\section{Introduction}
Recently, considerable attention has been focused on the problem of impurities 
embedded in a quantum chain. Using simpler bosonization and renormalization 
group techniques, Kane and Fisher have shown that a potential scattering center 
in a Luttinger liquid is driven to a strong-coupling fixed point by the 
repulsive interaction in the bulk\cite{1}. This is the first time to show that
 a single impurity in a one-dimensional quantum system behaves rather 
 different from that in a Fermi liquid, and directly stimulates the study on 
 the problem of local perturbations to a Luttinger liquid and especially on 
 the Kondo problem in a Luttinger liquid\cite{2,3,4,5}. It is well known now 
 that the spin dynamics of the Kondo problem is equivalent to that of a spin
  chain with an impurity\cite{6}.
\par
The integrable impurity problem of the Heisenberg chain with periodic boundary 
condition was first considered by Andrei and Johannesson\cite{7}. They studied 
the integrable case of a spin $S>1/2$ embedded in a spin-$1/2$ Heisenberg chain.
 Subsequently, the problem was generalized to arbitrary spins by Lee and 
 Schlottmann, and Schlottmann\cite{8}. Now the quantum inverse scattering
  method becomes a standard method to construct integrable impurity 
  models\cite{9}. A recent example is the integrable impurity problem in the 
  supersymmetric $t-J$ model\cite{10}. In the QISM, the hamiltonian of the 
  model is usually written as the logarithmic derivative of an homogeneous 
  transfer matrix at a special value of the spectral parameter. Inputting some
   inhomogeneous vertex matrices (provided these matrices satisfy the same 
   Yang-Baxter relation of the homogeneous ones), then we obtain an 
   inhomogeneous transfer matrix. Its logarithmic derivative at some special 
   value of the spectral parameter thus gives a hamiltonian with local 
   interactions. However, to construct an integrable impurity model with 
   periodic boundary condition, there is a prize to pay, namely some 
   unphysical terms must present in the hamiltonian, though they may be
    irrelevant. Their presence is required by the integrablility.
\par
In another hand, the open boundary problem for the quantum chain has been
 renewed due to Kane and Fisher's observation\cite{1}. It is found that the
  open boundary theory is very useful to formulate both the thermodynamics 
  and the transport properties of the quantum chains with 
  impurity\cite{11,12,13}. The open boundary problem of the integrable 
  models was first considered by Gaudin\cite{14}, who studied the nonlinear 
  Schr{\"o}dinger model and the spin-$1/2$ Heisenberg chain with simple open
   boundaries. Subsequently, his method was generalized to the Hubbard model
    by Schulz\cite{15} and the spin chain with boundary fields by Alcaraz et al.
    \cite{16}. In addition, Sklyannin formulated the same result based on the 
    QISM. In his theory, a new relation which now is called as the reflecting 
    Yang-Baxter relation\cite{17} was used.
\par
In this paper, we study the problem of an open spin$1/2$ Heisenberg chain 
coupled with two impurity spins sited at the ends of the system. The 
hamiltonian we shall consider reads
\begin{eqnarray}
H=\frac12J\sum_{j=1}^{N-1}{\vec \sigma}_j\cdot{\vec \sigma}_{j+1}+J_R{\vec 
\sigma}_N\cdot{\vec S}_R+J_L{\vec \sigma}_1\cdot{\vec S}_L,
\end{eqnarray}
where ${\vec \sigma}_j$ are the Pauli matrices; ${\vec S}_{R,L}$ are the
 impurity moments with an arbitrary spin $S$; $N$ is the site number of the 
 bulk; $J$ is a positive constant and $J_{R,L}$ are two arbitrary real 
 constants which describe the coupling between the bulk and the impurities.
  We remark that the present model is very reasonable for the absence of 
  redundant terms which can not be justified on physical grounds.
\par
The structure of the present paper is the following: In the subsequent section,
 we construct the transfer matrix corresponding to the hamiltonian (1), thus the
  integrability of the present model can be directly justified. Based on the 
  QISM, the Bethe ansatz equation and the eigenvalue of the hamiltonian will 
  be derived. In sect.$III$, the ground state properties for different regions 
  of parameter $J_{R,L}$ will be discussed. Sect.$IV$ is attributed to the
   residual entropy of the ground state and the low temperature specific heat. 
   Concluding remarks will be given in sect.$V$.
\section{Algebraic Bethe Ansatz}
In the framework of QISM, the integrable hamiltonian with open boundary 
condition is usually obtained from a monodromy matrix\cite{17}
\begin{eqnarray}
U(\lambda)=K_-(\lambda)T(\lambda)K_+(\lambda)T^{-1}(-\lambda),
\end{eqnarray}
where $K_\pm(\lambda)$ are the reflecting matrices which satisfy the 
reflecting Yang-Baxter relation
\begin{eqnarray}
S_{12}(\lambda-\mu)K_{\pm}^1(\lambda)S_{12}(\lambda+\mu)K_{\pm}^2(\mu)=
K_{\pm}^2(\mu)S_{12}(\lambda+\mu)K_{\pm}^1(\lambda)S_{12}(\lambda-\mu),
\end{eqnarray}
with $K_\pm^{1,2}$ and $S_{12}$ acting on the space $V_{1,2}$ and $V_1\otimes 
V_2$ respectively. $S_{12}$ is the scattering matrix which satisfy the
 traditional Yang-Baxter relation\cite{18}. $T(\lambda)$ is the monodromy
  matrix for the periodic system which satisfy the Yang-Baxter relation
\begin{eqnarray}
S_{12}(\lambda-\mu)T_1(\lambda)T_2(\mu)=T_2(\mu)T_1(\lambda)S_{12}(\lambda-\mu).
\end{eqnarray}
As demonstrated by Sklyannin\cite{17}, $U(\lambda)$ satisfies the same
 reflecting Yang-Baxter relation (3) as $K_{\pm}$ do. Thus the trace of 
 $U(\lambda)$ gives an infinite number of conserved quantities. We remark 
 that Sklyannin and many other authors used c-number $K_\pm(\lambda)$ to 
 construct their models, where $K_\pm(\lambda)$ only induce the boundary 
 fields. The operator ones were first used by the present author and 
 coworkers\cite{5} to study the Kondo problem in one-dimensional strongly
  correlated electron systems.
\par
To construct the algebraic Bethe ansatz of the present model, we define
\begin{eqnarray}
T_{\tau}(\lambda)=L_{R\tau}^+(\lambda)L_{N\tau}(\lambda)L_{N-1\tau}(\lambda)
\cdots L_{1\tau}(\lambda),\nonumber\\
{\tilde T}_{\tau}(\lambda)=L_{1\tau}(\lambda)L_{2\tau}(\lambda)\cdots L_{N\tau}
(\lambda)L_{R\tau}^-(\lambda),
\end{eqnarray}
where $L_{j\tau}(\lambda)=i\lambda+1/2(1+{\vec \sigma}_j\cdot{\vec \tau})$, 
and $L_{R\tau}^\pm(\lambda)=i\lambda\pm c_R+\frac12+{\vec \tau}\cdot{\vec S_R}$
 with ${\vec \tau}$ an auxiliary Pauli matrix. Furthermore, we put 
\begin{eqnarray}
K_-(\lambda)=1,\nonumber\\
K_+(\lambda)=(i\lambda+c_L+\frac12+{\vec \tau}\cdot{\vec S_L})(i\lambda-c_L+
\frac12+{\vec \tau}\cdot{\vec S_L}),
\end{eqnarray}
where $c_{R,L}$ are two  constants. Obviously, $K_\pm(\lambda)$ defined in 
(6) satisfy the reflecting relation (3). The monodromy matrix $U(\lambda)$
 is defined for our model as 
\begin{eqnarray}
U_\tau(\lambda)=K_-(\lambda)T_\tau(\lambda)K_+(\lambda){\tilde T}_\tau(\lambda),
\end{eqnarray}
which satisfy the reflecting equation
\begin{eqnarray}
S_{12}(\lambda-\mu)U_{\tau_1}(\lambda)S_{12}(\lambda+\mu)U_{\tau_2}(\mu)=
U_{\tau_2}(\mu)S_{12}(\lambda+\mu)U_{\tau_1}(\lambda)S_{12}(\lambda-\mu),
\end{eqnarray}
with $S_{12}(\lambda\pm\mu)=L_{\tau_1\tau_2}(\lambda\pm\mu)$. The hamiltonian
 (1) is obtained by the following relation
\begin{eqnarray}
H=\frac {-iJ}{4\prod_{r=R,L}[(S+\frac12)^2-c_r^2]}\frac{dX(\lambda)}
{d\lambda}|_{\lambda=0}-\frac N2J-\frac12\sum_{r=R,L}\frac J{(S+\frac12)^2-
c_r^2},
\end{eqnarray}
where $X(\lambda)=tr_\tau U_\tau(\lambda)$ and  $J_{R,L}$ are parametrized by 
$c_{R,L}$ as $J_{R,L}=J/[(S+1/2)^2-c_{R,L}^2]$.
\par
Although the model can be solved for arbitrary $J_{R,L}$, we consider only 
the $c_R=c_L=c$ ($J_R=J_L=J_i$) case in this paper. The general case can be 
formulated following the same procedure without any difficulty. We introduce
 the notation
\[ U_\tau(\lambda)=\left(\begin{array}{clcr}
A(\lambda) &B(\lambda)\\
C(\lambda) &D(\lambda)
\end{array}\right)\].
Some useful commutation relation can be formulated from (8) as
\begin{eqnarray}
{[A(\lambda),A(\mu)]=[A(\lambda),D(\mu)]=[B(\lambda),B(\mu)]}=0,\nonumber\\
A(\lambda)B(\mu)=\frac{(\lambda+\mu)(\lambda-\mu+i)}{(\lambda-\mu)(\lambda+
\mu-i)}B(\mu)A(\lambda)-\frac{2i\mu}{(\lambda-\mu)(2\mu-i)}B(\lambda)A(\mu)
\nonumber\\
+\frac i{(\lambda+\mu-i)(2\mu-i)}B(\lambda){\bar D}(\mu),\\
{\bar D}(\lambda)B(\mu)=\frac{(\lambda-\mu-i)(\lambda+\mu-2i)}{(\lambda-\mu)
(\lambda+\mu-i)}B(\mu){\bar D}(\lambda) +\frac{2i(\lambda-i)}{(\lambda-\mu)
(2\mu-i)}B(\lambda){\bar D}(\mu) \nonumber\\
-\frac{4i(\lambda-i)\mu}{(2\mu-i)(\lambda+\mu-i)}B(\lambda)A(\mu),\nonumber
\end{eqnarray}
where ${\bar D}(\lambda)$ is defined as ${\bar D}(\lambda)=(2\lambda-i)
D(\lambda)+iA(\lambda)$. Therefore, the trace of $U_\tau(\lambda)$ can be 
expressed as
\begin{eqnarray}
X(\lambda)\equiv Tr_\tau U_\tau(\lambda)=\frac 1{2\lambda-i}{\bar D}(\lambda)+
\frac{2\lambda-2i}{2\lambda-i}A(\lambda).
\end{eqnarray}
Define the pseudo vacuum state $|0>$ as
\begin{eqnarray}
\sigma_j^+|0>=S^+|0>=0.
\end{eqnarray}
The elements of $U_\tau(\lambda)$ acting on the pseudo vacuum state behave as
\begin{eqnarray}
C(\lambda)|0>=0,\nonumber\\
A(\lambda)|0>=a(\lambda)|0>=[(i\lambda+S+\frac12)^2-c^2]^2(i\lambda+1)^{2N}|0>,\\ 
{\bar D}(\lambda)|0>={\bar d}(\lambda)|0>=(i\lambda)^{2N}[(i\lambda-S+\frac12)^2-c^2]^2|0>.\nonumber
\end{eqnarray}
Therefore, $X(\lambda)$ can be treated as the generating functional of an 
infinite number of conserved quantities (including the hamiltonian) and
 $B(\lambda)$ is the creation operator of their eigenstates. An eigenstate
  of $X(\lambda)$ with $M$ spins down can be constructed as
\begin{eqnarray}
|\Omega>=\prod_{j=1}^MB(\lambda_j)|0>.
\end{eqnarray}
With the relations (10) and (13), we obtain the eigenvalue of $X(\lambda)$ 
acting on the state $|\Omega>$ as
\begin{eqnarray}
X(\lambda)|\Omega>=\Lambda(\lambda;\lambda_1,\cdots,\lambda_M)|\Omega>,
\nonumber\\
\Lambda(\lambda;\lambda_1,\cdots,\lambda_M)=\frac {{\bar d}(\lambda)}
{2\lambda-i}\prod_{j=1}^M\frac{(\lambda-\lambda_j-i)(\lambda+\lambda_j-2i)}
{(\lambda-\lambda_j)(\lambda+\lambda_j-i)}\\
+\frac{2(\lambda-i)a(\lambda)}{2\lambda-i}\prod_{j=1}^M\frac{(\lambda+\lambda_j)
(\lambda-\lambda_j+i)}{(\lambda+\lambda_j-i)(\lambda-\lambda_j)}.\nonumber
\end{eqnarray}
However, the spectral parameters $\lambda_j$ are not independent each other but
 satisfy the following Bethe ansatz equation
\begin{eqnarray}
\big(\frac{\lambda_j+\frac i2}{\lambda_j-\frac i2}\big)^{2N}\big(\frac
{\lambda_j-ic+iS}{\lambda_j-ic-iS}\big)^2\big(\frac{\lambda_j+ic+iS}
{\lambda_j+ic-iS}\big)^2=\prod_{r=\pm1}\prod_{l\neq j}^M\frac{\lambda_j-r\lambda_l+i}{\lambda_j-r\lambda_l-i}.
\end{eqnarray}
The eigenvalue of the hamiltonian (1) acting on the state $|\Omega>$ 
can be obtained from (9) and (15) as
\begin{eqnarray}
E(\lambda_1,\cdots,\lambda_M)=-\sum_{j=1}^M\frac J{\lambda_j^2+\frac14}+
\frac12(N-1)J+2SJ_i.
\end{eqnarray}
\section{ground state}
Since the parameter $J_i$ depends only on $c^2$, $c$ may take either real 
or  imaginary values. In this section, we discuss the ground state properties
 of different $c$ values.
\par
(i). $c>S+1/2$ case. In this case, the coupling between the bulk and the 
impurity is ferromagnetic. All 
the $\lambda$ modes take real values in the ground state to minimize the 
energy. Taking the logarithm of (16) we obtain
\begin{eqnarray}
\frac{I_j}N=\frac 1\pi\{\theta_1(\lambda_j)+\frac 1{2N}[\phi(\lambda_j)-
\sum_{l=-M}^M\theta_2(\lambda_j-\lambda_l)]\},
\end{eqnarray}
where $\theta_n(\lambda)=2\tan^{-1}(2\lambda/n)$, $\phi(\lambda)=
2\theta_{2(S+|c|)}(\lambda)-2\theta_{2(|c|-S)}(\lambda)+
\theta_2(\lambda)+\theta_1(\lambda)$, $I_j$ are some integers. 
Note above we have included the negative modes by putting $\lambda_j
=-\lambda_{-j}$ ($\lambda_0=0$). Define
\begin{eqnarray}
Z_N(\lambda)=\frac 1\pi\{\theta_1(\lambda)+\frac 1{2N}[\phi(\lambda)-
\sum_{l=-M}^M\theta_2(\lambda-\lambda_l)]\}.
\end{eqnarray}
Then $Z_N(\lambda_j)=I_j/N$ gives the Bethe ansatz equation (18). For the 
ground state, $I_j$ must take consecutive numbers around zero symmetrically.
 A density function for the ground state in the thermodynamic limit can be 
 defined as
\begin{eqnarray}
\rho_N(\lambda)=\frac{dZ_N(\lambda)}{d\lambda},
\end{eqnarray}
with the condition $\int_{-\Lambda}^\Lambda\rho_N(\lambda)d\lambda=(2M+1)/N$,
 where $\Lambda$ is the cutoff of $\lambda$ modes. As discussed in many earlier
  papers\cite{19}, the eigenenergy is minimized at $\Lambda=\infty$ up to the 
  order $O(N^{-2})$. With this condition, we obtain that $M=N/2$, which gives 
  the self magnetization of the ground state as
\begin{eqnarray}
M_g=\frac 12N-M+2S=2S.
\end{eqnarray}
Such a result indicates that the impurity moments can not be screened due to
 the ferromagnetic coupling between the bulk and the impurity.
\par
(ii). $S<|c|<S+1/2$. In this case, $J_i>0$ and thus the exchange interaction 
between the impurity and the bulk falls into the antiferromagnetic regime. 
Two imaginary modes of $\lambda$ at $\lambda=i(|c|-S)$ can exist in the ground
 state. Note this mode carries energy $\epsilon_i=-1/[1/4-(|c|-S)^2]$, which 
 is smaller than those carried by any real modes. The real modes thus satisfy 
 the following Bethe ansatz equation
\begin{eqnarray}
\big(\frac{\lambda_j+\frac i2}{\lambda_j-\frac i2}\big)^{2N}\big(
\frac{\lambda_j-ic+iS}{\lambda_j-ic-iS}\big)^2\big(\frac{\lambda_j+ic+iS}
{\lambda_j+ic-iS}\big)^2=\big(\frac{\lambda_j-i|c|+iS+i}{\lambda_j-i|c|+iS-i}
\big)^2\big(\frac{\lambda_j+i|c|-iS+i}{\lambda_j+i|c|-iS-i}\big)^2\nonumber\\
\times\prod_{r=\pm1}\prod_{l\neq j}^{M-2}\frac{\lambda_j-r\lambda_l+i}
{\lambda_j-r\lambda_l-i}.
\end{eqnarray}
With the same procedure discussed in case (i), we obtain $M=(N+2)/2$ or the 
residual magnetization of the ground state $M_g=2S-1$ as expected. This means
 the impurity moment is partially screened, a similar result to that of the 
 Kondo problem\cite{6}.
\par
(iii). $0<|c|<S$. In this case, the system falls also into the regime of 
antiferromagnetic coupling ($J_i>0$). No bound state can exist in the ground
 state. The function $Z_N(\lambda)$ can be defined in a similar way of the
  case(i) but with a different $\phi(\lambda)=2\theta_{2(S+c)}(\lambda)+
  2\theta_{2(S-c)}(\lambda)+\theta_2(\lambda)+\theta_1(\lambda)$. With the
   condition $Z_N(\pm\infty)=\pm(M+1/2)/N$ we have again $M=(N+2)/2$ and 
   the residual magnetization $M_g=2S-1$.
\par
(iv). When $c$ takes an imaginary value, $J_i$ is always positive. Suppose 
$c=ib$. The Bethe ansatz equation (15) then becomes
\begin{eqnarray}
\big(\frac{\lambda_j+\frac i2}{\lambda_j-\frac i2}\big)^{2N}
\big(\frac{\lambda_j-b+iS}{\lambda_j-b-iS}\big)^2\big(\frac{\lambda_j+b+iS}{\lambda_j+b-iS}\big)^2=\prod_{r=\pm1}\prod_{l\neq j}^M\frac{\lambda_j-r\lambda_l+i}{\lambda_j-r\lambda_l-i}.
\end{eqnarray}
For the ground state, all $\lambda$ take real values and their cutoff $\Lambda$
 still tends to infinity in the thermodynamic limit. As discussed in case (i), 
 a similar function $Z_N(\lambda)$ can be defined but with a different
  $\phi(\lambda)=2\theta_{2S}(\lambda-b)+2\theta_{2S}(\lambda+b)+
  \theta_2(\lambda)+\theta_1(\lambda)$. Therefore $M=(N+2)/2$ for the
   ground state, which still leaves a residual magnetization of $2S-1$.
\par
From the above discussion we conclude that the impurities can be screened 
(partially for $S>1/2$) only in the case where $J_i>0$. The impurity moments
 can not be screened when $J_i<0$, unlike the situation for the Kondo problem
  in a Luttinger liquid predicted by Furusaki and Nagaosa\cite{3}.
\section{residual entropy and low temperature specific heat}
The thermodynamics of the present model can be constructed with the standard
 method proposed by Yang and Yang\cite{20} based on the string 
 hypothesis\cite{21}. Here we omit the details which can be found 
 in some nice reviews\cite{22}. An interesting feature of our model 
 is that the residual entropy $S_g$ may take different values depending
  on the bond deformation between the bulk and the impurities:
\[S_g=\left \{\begin{array}{llll}
\ln\frac{[2|c|]+2S}{[2|c|]-2S},&\mbox{if $|c|>S+\frac12$},\\
\ln\frac{(2S+[2|c|])(2+[2|c|]-2S)}{([2|c|]-2S)(2-[2|c|]+2S)},
&\mbox{if $S+\frac12>|c|>S$},\\
2\ln\sqrt{4S^2-[2|c|]^2},&\mbox{if $S>|c|>0$}\\
2\ln(2S),&\mbox{if c imaginary},
\end{array}
\right. \]
where $[2|c|]$ is the maximum integer equal or less than $2|c|$.  
For $|c|<1/2$ or imaginary and $S=1/2$, $S_g=0$, which means the 
impurity spin can be completely screened in the ground state and thus 
the system flows to a local Fermi liquid fixed point at low energy scales. 
The low temperature thermodynamics can be formulated based on the local
 Fermi liquid theory for the Kondo problem\cite{23}. Since the spectrum 
 are described by the quantum numbers $I_j$, $p_N(\lambda_j)=
 \pi Z_N(\lambda_j)$ can be treated as the momenta of the quasi-particles 
 in the Luttinger-Fermi liquid picture for the integrable models\cite{24}. 
 In the thermodynamic limit, the ground state energy can be expressed as
\begin{eqnarray}
E_g=\frac12N\int\epsilon_0(\lambda)\rho_N(\lambda)d\lambda+const.,
\end{eqnarray}
up to the order of $O(N^{-2})$, where $\epsilon_0(\lambda)=-J/(\lambda^2+1/4)$. From the definition of $\rho_N(\lambda)$ in (20), we can rewrite (24) as
\begin{eqnarray}
E_g=\frac N{4\pi}\int\epsilon(\lambda)\rho_N^{(0)}(\lambda)d\lambda+const.,
\end{eqnarray}
where
$\epsilon(\lambda)=\frac{-\pi J}{\cosh(\pi\lambda)}$ is
the dressed energy function, which can be treated as the energy of the 
``quasi-particles". Note that $\rho_N(\lambda)$ for $c$ imaginary can be 
solved up to $O(N^{-1})$ as
\begin{eqnarray}
\rho_N(\lambda)=\rho_0(\lambda)+\frac1N\rho_i(\lambda)+\frac1N\rho_b(\lambda),\nonumber\\
\rho_0(\lambda)=\frac1{\cosh(\pi\lambda)},\\
\rho_b(\lambda)=\frac1{2\cosh(\pi\lambda)}+\int\frac{e^{-\frac12|\omega|}e^{-i\lambda\omega}}{4\pi\cosh(\frac12\omega)}d\omega,\nonumber\\
\rho_i(\lambda)=\frac1{\cosh(\lambda-b)}+\frac1{\cosh(\lambda+b)},\nonumber
\end{eqnarray}
where $\rho_0(\lambda)$, $\rho_i(\lambda)/N$, and $\rho_b(\lambda)/N$ are the 
contributions of the bulk, the impurity and the open boundary to the density 
respectively. The density of states at the Fermi surface can be determined in 
the Fermi liquid picture as
\begin{eqnarray}
N(0)=\frac1{2\pi}\frac{dp_N(\lambda)}{d\epsilon(\lambda)}|_{\lambda=\infty}
=\frac{\rho_N(\lambda)}{2\epsilon'(\lambda)}|_{\lambda=\infty}
\end{eqnarray}
Therefore, the impurity contribution to the low temperature specific heat 
reads
\begin{eqnarray}
C_i=\frac{2\pi}{3NJ}\cosh(\pi b)T,
\end{eqnarray}
While for $|c|<1/2$, $\rho_i(\lambda)$ can be solved by a similar way as
\begin{eqnarray}
\rho_i(\lambda)=\frac1{\cosh(\lambda-ic)}+\frac1{\cosh(\lambda+ic)}.
\end{eqnarray}
Thus the impurity specific heat reads
\begin{eqnarray}
C_i=\frac\pi{3NJ}\cos(\pi c).
\end{eqnarray}
Notice that the Kondo temperature is nothing but the effective Fermi 
temperature in the local Fermi liquid theory. Therefore, we conclude 
that the Kondo temperature for the cases discussed above is given by
\begin{eqnarray}
T_k=\pi J\cos^{-1}(\pi c).
\end{eqnarray}
Such a result directly shows the crossover from an exponential law to an 
algebraic one when $c$ goes from imaginary to real as pointed out earlier 
by Lee and Toner\cite{2}. Notice that $c=\sqrt{1-J/J_i}$ and for the Kondo
 problem in a Hubbard chain, the band width  of the spinons (proportional
  to $J$) is about $4t^2/U$. Where $2t$ is the band width of the fermions,
   $U$ is the onsite Coulomb repulsion. For the strong correlation limit 
   $U>>t$, $J\leq J_i$, the Kondo temperature shows a power law of 
   $t^2/(UJ_i)$. While for the weak correlation limit $U<<t$, 
   $J\geq J_i$, the Kondo temperature shows a exponential law 
   of $t^2/UJ_i$.  
\section{Concluding remarks}
In conclusion, we establish an exactly solvable model of Heisenberg chain 
coupled with two impurity moments. This model is relevant to the Kondo 
problem in a Luttinger liquid. With the algebraic Bethe ansatz method, 
the hamiltonian is exactly diagonalized. It is found that the local moments 
can be screened (partially for $S>1/2$) only in the case where the coupling 
between the bulk and the impurities is antiferromagnetic ($J_i>0$). Such a 
result strongly contradicts to that of the classical system (Ising model), 
where the ground state is a pure Neel state with total magnetization of zero
 (provided $N$ even) for both $J_i>0$ and $J_i<0$ and arbitrary $S$. The 
 present result shows that the quantum fluctuation plays a  crucial rule 
 in the one-dimensional quantum system. The residual entropy of the ground 
 state is derived from the thermal Bethe ansatz. It strongly depends on the 
 parameter $c$, a similar result to that of the Kondo problem in the Kondo 
 problem\cite{5}. Based on the local Fermi liquid theory\cite{23} and the 
 Landau-Luttinger description for the integrable models\cite{24}, we derive 
 the low temperature specific heat for some special cases. The Kondo
  temperatures are exactly carried out for these cases. The exact result 
  does show the crossover of the Kondo temperature from an exponential 
  law to a power law when the parameter $c$ goes from imaginary to real, 
  a phenomenon first obtained by Lee and Toner\cite{2}.\\ 
{\bf Acknoledgement}\\
This work was partially supported by the National Natural Science 
Foundation of China and the Natural Science Foundation for Young 
Scientists, Chinese academy of Sciences.

\end{document}